\documentclass[reprint, amsmath, amssymb, aps]{revtex4-1}

\usepackage{braket}
\usepackage{graphicx}% Include figure files
\usepackage{dcolumn}% Align table columns on decimal point
\usepackage{bm}% bold math
\usepackage{xcolor}% bold math
\usepackage[normalem]{ulem}
\newcommand{\Nu}{N_{\uparrow}}
\newcommand{\NuO}{N_{\uparrow p}}
\newcommand{\NuT}{N_{\uparrow f}}

\newcommand{\NdO}{N_{\downarrow p}}
\newcommand{\NdT}{N_{\downarrow f}}
\newcommand{\Jzp}{J_{zp}}
\newcommand{\Jzf}{J_{zf}}
\newcommand{\up}{\ket{\uparrow}}
\newcommand{\down}{\ket{\downarrow}}

\newcommand{\tfb}{t_{fb}}
\newcommand{\Jztar}{J_{ztar}}
\newcommand{\thetafb}{\theta_{fb}}
\newcommand{\Nproj}{N_{\downarrow QPN}}
\newcommand{\Jzproj}{J_{z,QPN}}

\newcommand{\Qsensitivity}{Q_{PD}}
\newcommand{\Qelectronic}{Q_{elec}}
\newcommand{\Qdeadtime}{Q_{turnon}}

\newcommand{\Lsha}{L_{SHA}}
\newcommand{\Lstab}{L_{stab}}
\newcommand{\Lcav}{L_{cav}}

\newcommand{\Qarea}{\tilde{Q}_{1}}
\newcommand{\Qcompound}{Q_{1}}
\newcommand{\Qcompoundexpected}{Q_{1}^{(0)}}

\begin{document}

\title{Deterministic Squeezed States with Collective Measurements and Feedback}

\author{Kevin C. Cox}
\affiliation{JILA, NIST, and University of Colorado, 440 UCB, Boulder, CO  80309, USA}
\author{Graham P. Greve}
\affiliation{JILA, NIST, and University of Colorado, 440 UCB, Boulder, CO  80309, USA}
\author{Joshua M. Weiner}
\affiliation{JILA, NIST, and University of Colorado, 440 UCB, Boulder, CO  80309, USA}
\author{James K. Thompson}
\affiliation{JILA, NIST, and University of Colorado, 440 UCB, Boulder, CO  80309, USA}

\date{\today}

\begin{abstract}

We demonstrate the creation of entangled, spin-squeezed states using a collective, or joint, measurement and real-time feedback. The pseudospin state of an ensemble of $N= 5\times 10^4$ laser-cooled $^{87}$Rb atoms is deterministically driven to a specified population state with angular resolution  that is a factor of 5.5(8) [7.4(6)~dB] in variance below the standard quantum limit for unentangled atoms --- comparable to the best enhancements using only unitary evolution.  Without feedback, conditioning on the outcome of the joint premeasurement, we directly observe up to 59(8) times [17.7(6)~dB] improvement in quantum phase variance relative to the standard quantum limit for $N=4\times 10^5$ atoms.  This is one of the largest reported entanglement enhancements to date in any system.

\end{abstract}

\maketitle

Entanglement is a fundamental quantum resource, able to improve precision measurements and required for all quantum information science.  Advances in the creation, manipulation, and characterization of entanglement will be required to develop practical quantum computers, quantum simulators, and enhanced quantum sensors.  In particular, quantum sensors operate by attempting to estimate the total amount of phase that accumulates between two quantum states, typically forming a pseudospin-$1/2$ system.  When $N$ atoms are unentangled, the independent quantum projection or collapse of each atom's wave function fundamentally limits the sensor by creating a rms uncertainty $\Delta\theta_{\text{SQL}}=1/\sqrt{N}$~rad in the estimate of the quantum phase, the  standard quantum limit (SQL) \cite{Itano1993}. However, entanglement can be used to create correlations in the quantum collapse of the $N$ atoms  \cite{Wineland1992,Kitagawa1993} to achieve large enhancements in phase resolution, in principle down to the Heisenberg limit $\Delta\theta_{\text{HL}}=1/N$~rad.

This Letter features two main results.  First, following Fig. \ref{fig1}(a), we use the outcome of a collective, or joint, measurement to actively steer the collective spin projection of an ensemble of $5\times10^4$ laser-cooled and trapped $^{87}$Rb atoms to a target entangled quantum state.  Real-time feedback allows generation of the target state with enhanced angular resolution $S^{-1} \equiv (\Delta\theta_{\text{SQL}}/\Delta\theta)^2=5.5(8)$, or 7.4(6)~dB below the SQL, with no background subtractions.  Second, we perform a direct subtraction of quantum noise without feedback and directly observe a conditionally enhanced phase resolution  $S^{-1}= 59(8)$ or equivalently 17.7(6)~dB below the SQL. Along with another recent result using similar collective measurements \cite{Kasevich18dB}, this is the largest phase enhancement from entanglement to date in any system.

Entanglement is  often created and manipulated via unitary interactions between qubits 
\cite{WinelandNature2000,Leibfried2005,BlattSqueeze,Noguchi2012,a_klempt_squeezing,a_treutline_Squeezing, Chapman12,Oberthaler2014,Bucker2011,LSM10}. 
However, the joint measurements on two or more qubits used here (sometimes referred to as quantum nondemolition measurements) have shown promise for creating entanglement, particularly among large numbers of qubits \cite{a_kuzmichQND,AWO09,SLV10,Wasilewski2010,Leroux10,CBS11,Squeezing_Bohnet_2014,a_mitchell_singlet,SKN12,2015arXiv150602297N, a_DiCarlo_EntanglementAndFeedback,a_siddiqi_remoteEntanglement}.  By adding real-time feedback guided by the outcome of joint measurements, one can access a more diverse range of quantum technologies including Heisenberg-limited atomic sensors \cite{a_sorensen_FB}, reduction of mean field shifts in atom interferometers \cite{a_OberthalerNumberSqueezing,a_ketterle_BECINterferometer_Squeezing}, quantum teleportation \cite{a_Polzik_Teleportation, Polzik_Teleport_Exp}, and error correction \cite{a_colorcode_Blatt,a_Scholkopf100PhotonCats}.  Quantum noise suppression with real-time feedback has been considered theoretically \cite{a_wisemanFB_PRA,a_Wiseman_FB_JPB} and demonstrated in a previous experiment \cite{a_TakahashiFeedback} but without the critical enhancement in phase resolution that signifies entanglement.

\begin{figure}
\centering
\includegraphics[width=3.375in]{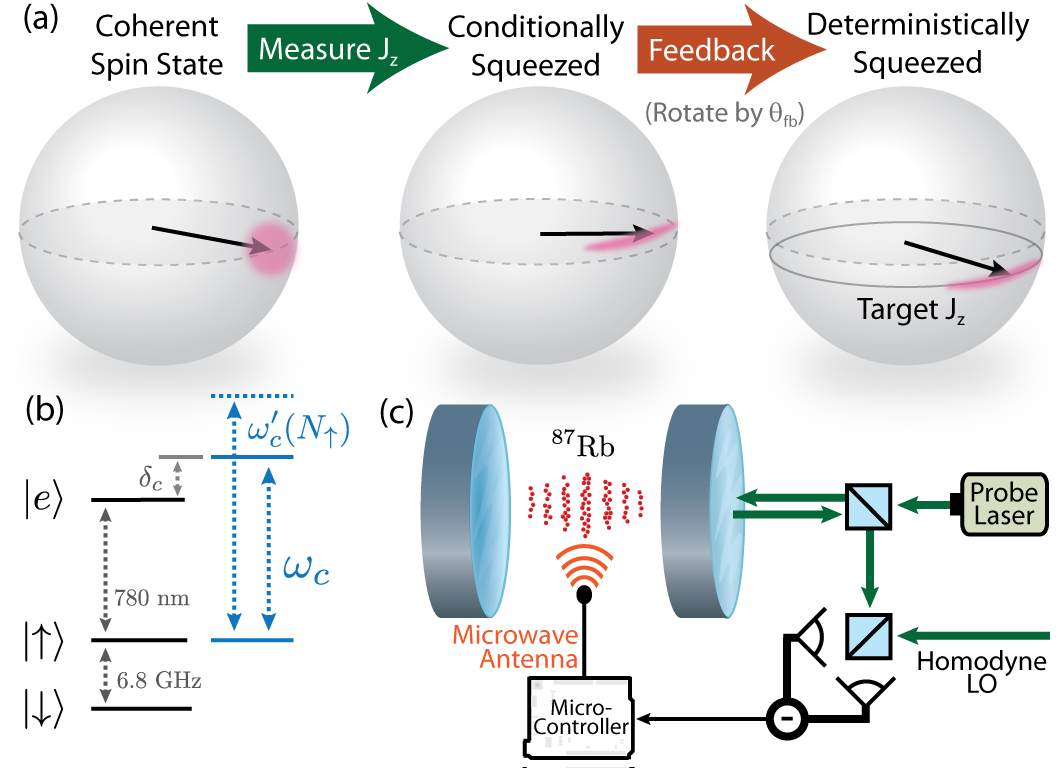}
\caption{ (a) A coherent spin state's spin-projection noise (pink distribution) is projected onto a squeezed state by a measurement of $J_z$.  The quantum state randomly collapses within the original distribution, creating a conditionally squeezed state. The premeasurement's outcome is then used to rotate the spin state's polar angle to a desired target spin projection (black solid line)  $J_z = \Jztar$, creating a deterministically squeezed state. (b) The relevant $^{87}$Rb energy levels (black) and cavity resonance frequency $\omega_{c}$ (blue).  (c) Simplified experimental diagram.  The cavity is probed in reflection.  Homodyne detection of the probe is sampled by a microcontroller that then applies microwaves at 6.8~GHz to achieve the desired feedback rotation $\thetafb$ to create the deterministically squeezed state in (a). See the Supplemental Material for experimental details \cite{Note1}. }
\label{fig1}
\end{figure}

We visualize a collection of $N$ spin-1/2 atoms as a single collective Bloch vector $\mathbf{J} = J_x \hat{x} + J_y \hat{y} + J_z \hat{z} $ given by first order expectation values $J_\alpha \equiv \braket{ \hat{J}_\alpha }$ of collective spin-projection operators with $\alpha = \{x,y,z\}$.  The quantum projection noise (QPN) and resulting SQL can be intuitively visualized by a quasiprobability distribution perpendicular to the classical Bloch vector [Fig. \ref{fig1}(a)]. The distribution's rms fluctuations along a given spin-projection direction are given by  $\Delta J_\alpha \equiv \sqrt{\braket{\hat{J}_\alpha^2}-\braket{\hat{J}_\alpha}^2}$.  In this Letter $\Delta$ will refer to the standard deviation of a given quantity.  For a coherent spin state oriented at the equator of the Bloch sphere, the spin projection $J_z$  and spin population $N_{\downarrow}$  both fluctuate from one trial to the next with a standard deviation $\Delta J_{z,QPN}= \Delta \Nproj \equiv \sqrt{N}/2$.   

We calculate the enhancement in phase resolution $S \equiv (\Delta \theta / \Delta \theta_{\text{SQL}})^2 = R /C^2$ \cite{Wineland1992}, where  $R \equiv (\Delta J_z/\Delta J_{z,QPN})^2$ is the observed spin-projection noise relative to the projection noise level, and $C \equiv 2 \braket{|\hat{J}|}/N$ is the fractional atomic coherence remaining (or ``contrast'') after a joint measurement.  An additional 0.2~dB correction is applied to $S$ for a 4\% background loss of contrast (see the Supplemental Material \cite{Note1}).  Observing $S^{-1}>1$ serves as a witness for entanglement between atoms \cite{Sorensen_EntanglemetnWitness} and the  magnitude usefully quantifies the degree of entanglement \cite{Wineland1992,Kitagawa1993}. 

A joint measurement of the population of atoms $N_\uparrow$ is engineered by measuring the frequency shift of a TEM$_{00}$ cavity mode.  The cavity is tuned $\delta_c = 2\pi \times 400$~MHz  to the blue of the $^{87}$Rb $\ket{\uparrow} \equiv \ket{5^2S_{1/2},F=2,M_F=2}$ to $\ket{e} \equiv \ket{5^2P_{3/2},F=3,M_F = 3}$ optical atomic transition as shown in Fig.~\ref{fig1}(b). The second state forming the pseudo-spin system is $\ket{\downarrow} \equiv \ket{5^2S_{1/2},F=1,M_F=1}$. The cavity has finesse $2532(80)$  and power decay linewidth $\kappa = 2\pi\times 3.15(10)$~MHz. The atoms are laser cooled to $10~\mu$K and trapped tightly on axis in an intracavity 1D optical lattice [Fig.~\ref{fig1}(c)]. Spatially inhomogeneous coupling of atoms to the cavity mode is handled as in Refs.~\cite{CBS11, a_SSPRA, Squeezing_Bohnet_2014,2015arXiv151001021H}.  Atoms in  $\ket{\uparrow}$ strongly phase shift the intracavity probe light, causing the empty cavity resonance frequency $\omega_c$ to shift to $\omega_c'$.  A measurement of the shift $\omega_c'-\omega_c$ using homodyne detection of probe light reflected from the cavity can then be used to infer the population $N_\uparrow$.   To measure the population $N_\downarrow$, a $\pi$-pulse microwave coupling can then be applied to swap the populations between $\up$ and $\down$, and a measurement of the new population in $\up$ can be made with the measurement outcome now labeled $N_\downarrow$.

\begin{figure}[htb]
\centering
\includegraphics[width=3.375in]{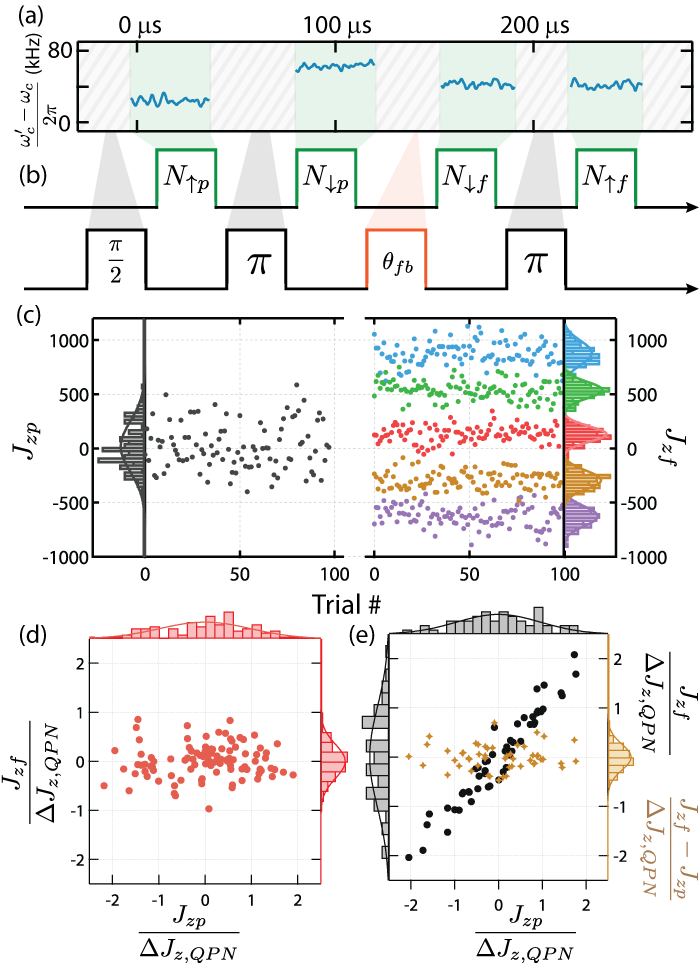}
\caption{(a) Measured cavity resonance frequency for a single trial versus time, subtracting a constant 12 MHz frequency offset.  (b)  The time windows in which the probe is turned on (green) and the populations determined from each window.  The fixed microwave rotations are shown in black with the feedback rotation is shown in orange.  (c) The premeasurements $\Jzp$ (left) and final measurements $\Jzf$  (right) of $J_z$ are plotted versus trial number and accumulated into histograms.  Five different $J_z$ states are targeted (five distinct colors on the right) and reached with noise below the QPN.  The maximum deterministic squeezing is $S = -7.4(6)$~dB relative to the SQL.  (d) Feedback reduces the noise distribution of the final measurement relative to the initial quantum noise in the premeasurement. (e) If no feedback is applied the final measurement and premeasurement are strongly correlated (black), allowing for conditional squeezing [$S = -10.3(6)$ dB] by using the differential quantity $\Jzf-\Jzp$ (gold).  The increase in noise from feedback is discussed in the Supplemental Material \cite{Note1}.}
\label{fig2}
\end{figure}

The experimental sequence is shown in Figs.~\ref{fig2} (a) and~\ref{fig2}(b). All atoms are prepared in $\down$, then a microwave $\pi/2$ pulse is applied to place each atom in an equal superposition of spin states, equivalent to preparing the Bloch vector along $\hat{y}$.  We make a measurement of the spin projection $J_z$ with measurement outcome labeled $\Jzp = (\NuO-\NdO)/2$.  Each population measurement outcome $\NuO$ and $\NdO$ is obtained by averaging the cavity-probe signal over a 40~$\mu$s window.  In each run of the experiment, a microcontroller calculates $\Jzp$ and applies feedback to steer the state toward a targeted value of spin projection $\Jztar$.  The feedback is accomplished by applying microwaves to rotate the Bloch vector through polar angle $\thetafb\approx 2\times(\Jztar-\Jzp)/(NC)$.  After the feedback, a final measurement of the spin projection $J_z$ is made with measurement outcome labeled $\Jzf=(\NuT-\NdT)/2$.
Feedback toward $\Jzf=0$ is evident in the time trace [Fig. \ref{fig2} (a)], since the  final two cavity frequency measurement windows that provide $\NuT$ and $\NdT$ are more nearly equal than was the case for the two premeasurement windows.

The microcontroller sets the sign of the rotation $\thetafb$ by digitally toggling between two microwave sources that are $180^\circ$ out of phase.  The magnitude of the rotation $\left| \thetafb\right|$ is controlled by varying the duration $\tfb$ for which the microwaves are applied, with a discrete timing resolution of approximately 12~ns.  The input technical noise floor, timing jitter, and timing resolution of the microcontroller are all sufficient to allow up to 20~dB of squeezing. 

The outcomes $\Jzp$ and ${\Jzf}$ are plotted versus trial number and collated into histograms in Fig.~\ref{fig2}(c).  Projection noise for this data (independently confirmed by measuring the scaling of $\Delta J_z$ with $N$) is $\Delta \Jzproj$ = 218(10), consistent with the measured $\Delta \Jzp = 235(24)$\color{black}. The data on the right shows the final measurement outcomes $\Jzf$ after applying feedback for five different target states $\Jztar$.  By implementing the feedback, each target state was reached with noise below the original projection noise.   

To observe deterministic squeezing or phase resolution enhancement, the atomic coherence that remains after the premeasurement and feedback must be evaluated.  The contrast is determined in a separate set of experiments by using microwave rotations after the feedback step to rotate the Bloch vector to determine its total length.  Accounting for the loss of coherence, we directly observe up to $S^{-1}= 5.5(8)$ [$7.4(6)$~dB] of deterministic squeezing via premeasurement and feedback.  
\begin{figure*}
\centering
\includegraphics[width=7in]{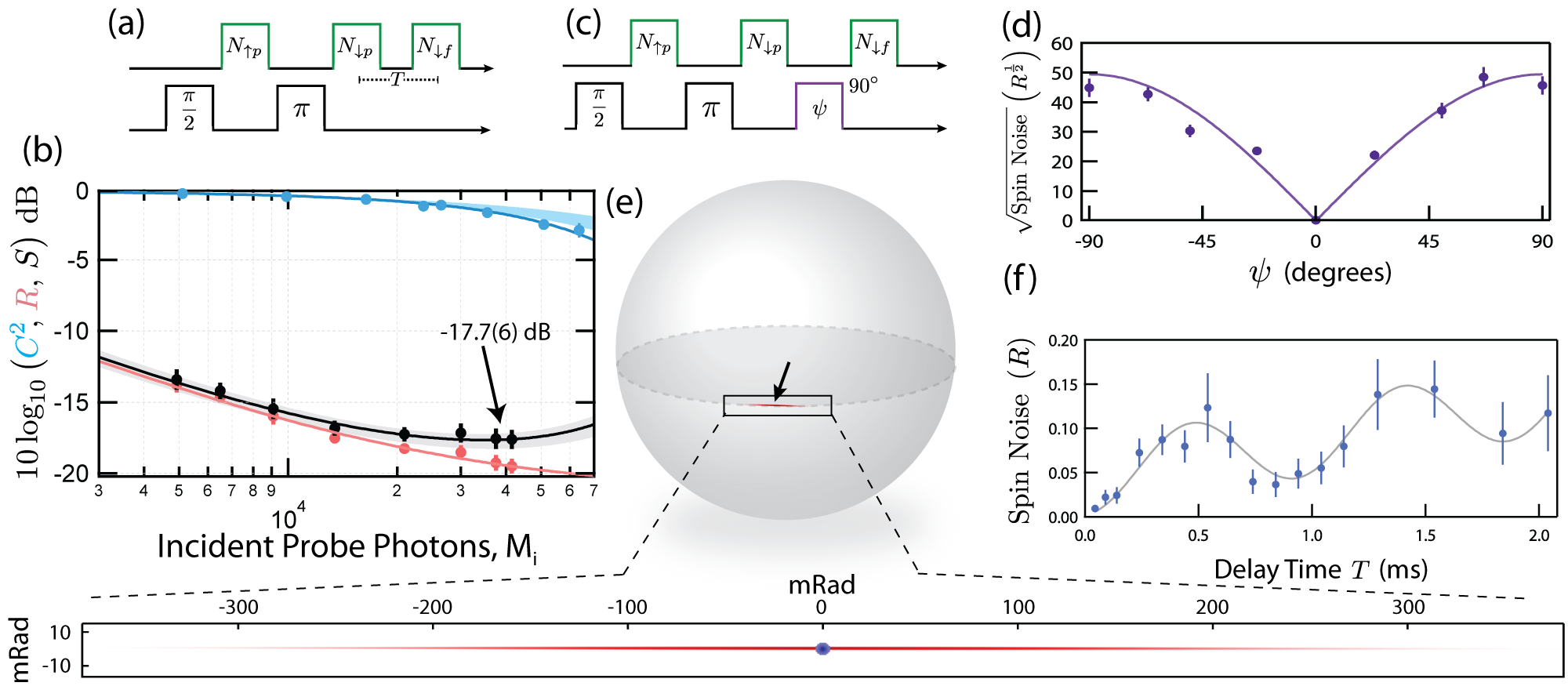}
\caption{
(a) Experimental sequence for conditional spin squeezing, with labeling mirroring that of Fig.~\ref{fig2}a.  (b) Squared contrast $C^2$ (blue), spin noise $R$ (red), and spin squeezing $S$ (black) are plotted versus the average number of incident photons $M_i$ in a single measurement window.  The solid lines are fits, the blue band is the predicted loss of contrast from free-space scattering, and the grey band indicates the total squeezing error bar.  (c) The experimental sequence used to observe the backaction spin projection.  (d) The measured spin noise $R$ is plotted versus $\psi$ with fit (purple).  (e) The reconstructed conditional probability distribution of the quantum state (red) on a Bloch sphere with Bloch (black) vector.  The distribution is magnified with a 1:1 aspect ratio and plotted with the equivalent coherent spin state (blue) in the lower panel.  (f) Thermal radial motion of the atoms causes the spin noise $R$ to oscillate at twice the radial trap frequency as the time separation $T$ between the pre- and final measurements is increased. }
\label{fig3}
\end{figure*}

For some applications, the feedback may not be necessary. Instead of applying feedback, one can cancel the quantum noise by directly subtracting the premeasurement  $\Jzp$ from the final measurement $\Jzf$, a technique known as conditional squeezing \cite{a_kuzmichQND,AWO09,SLV10,Wasilewski2010,Leroux10,CBS11,Squeezing_Bohnet_2014,a_mitchell_singlet,SKN12}.  In Figs.~\ref{fig2}(d) and~\ref{fig2}(e), we compare conditional and deterministic spin noise reductions taken under identical settings.  $\Jzf$ is plotted versus $\Jzp$ and the results are collated into histograms on each axis.  With feedback (red), $\Jzf$ is driven to zero with resolution below $\Delta \Jzproj$, regardless of $\Jzp$.  Without feedback (black), $\Jzp$ and $\Jzf$ are correlated, and the quantum noise can be conditionally subtracted from the final measurement by taking the difference $\Jzf-\Jzp$ (gold).

The deterministic squeezing with feedback is primarily limited by errors in the $\pi$ pulses due to microwave amplitude and frequency noise.  However, by increasing the number of atoms to $N=4\times10^5$, we improve the amount of conditional spin squeezing to $S^{-1}=59(8)$ or 17.7(6)~dB.  The experimental measurement sequence is the same, but to avoid added noise from the $\pi$ pulses, we only consider the reduction in the noise of the  difference of two population measurements of the same spin state $R =(\Delta(\NdT-\NdO))/\Delta N_{\downarrow \text{QPN}})^2$ [Fig.~\ref{fig3}(a)].  The information gained from the first measurement $\NuO$ is not used here, but its presence serves to spin echo away probe-induced inhomogeneous light shifts at the end of the premeasurement pair $\NuO$ and $\NdO$. Because the Bloch vector lies at the equator, small angular displacements of the polar angle could be sensed from changes in a single spin state's population alone. 

In Fig.~\ref{fig3}(b), we show the noise reduction $R$ versus the average number of photons $M_i$ incident upon the cavity during a single probe measurement window.  Again, this is the directly observed noise reduction with no background subtractions or removal of noise of the final measurement applied.  The maximum quantum noise reduction is $R^{-1} = 92(9)$, or $19.6(4)$~dB below the QPN and is limited by both a technical noise floor 25~dB below the QPN and optomechanical effects induced by the probe light being turned on and off, an effect that increases with $M_i$.  Also apparent in Fig.~\ref{fig3}(b), the atomic coherence or contrast (blue) after the premeasurement decreases with increasing $M_i$ due primarily to undesired free space scattering causing collapse of the individual atoms' wave functions into spin up (blue prediction band). The background contrast $C_{\text{BG}}$ is obtained from a measurement with $M_i=0$ in the two premeasurement windows. The black data and fit in Fig.~\ref{fig3}(b) display the squeezing obtained by combining the reduction in noise with the reduction in contrast.

We also examine the backaction or antisqueezed spin projection.  The experimental sequence is shown in Fig. \ref{fig3}(c) and is distinguished by the replacement of the rotation $\thetafb$ with a microwave rotation about an axis parallel to the Bloch vector through a fixed angle $\psi$.  Figure~\ref{fig3}(d) shows the increase in spin noise $R$ moving from the 17~dB squeezed (at $\psi=0$) to antisqueezed (at $\psi=\pm 90^\circ$) projections. Using an inverse Radon transform, we construct a visualization of the equivalent squeezed state, shown in Fig.~\ref{fig3}(e). The original coherent state noise is shown in blue. The state has $\Delta J_z \Delta J_x / (\Delta \Jzproj)^2 = 6.1 >1$ and is no longer a minimum uncertainty state owing to finite quantum efficiency for detecting the probe light.    From the increase in area and its scaling with $M_i$ we can infer the quantum efficiency of a joint measurement of a single population is $\tilde{Q}_{1}= 38(14)\%$, in good agreement with an independent prediction of 37(5)\% from measuring path efficiencies, cavity loss, detector efficiencies, technical noise floors, and laser turn-on times (see the Supplemental Material \cite{Note1}). Here, the total quantum efficiency of the full measurement sequence ($\NuO,\NdO,\NdT$) is effectively 4 times lower than $\tilde{Q}_{1}$ due to the additional noise in the final measurement $\NdT$ and the presently unused premeasurement $\NuO$.

In Fig. \ref{fig3}(f), we evaluate how well the conditional noise reduction can be maintained over a variable evolution time $T$.   This is an important consideration for implementing conditional squeezing in atomic sensors.  The contribution to $R$ from technical noise sources is partially removed by performing the measurement sequence of Fig.~\ref{fig3}(a) with no atoms present and subtracting the measured noise variance from the noise variances obtained with atoms present.  The spin noise $R$ is seen to oscillate at twice the radial frequency of the trapping potential due to thermal radial atomic motion that causes an oscillation in each atom's coupling to the cavity mode.  The additional monotonic increase in $R$ is not currently understood. A 3D optical lattice or a smaller atomic temperature to lattice depth ratio can be used to reduce the noise oscillations in the future.

The improved squeezing relative to previous work \cite{Squeezing_Bohnet_2014, DephasingPaper} was achieved by increasing the net quantum efficiency for probe detection from 5\% to 37\% (by constructing a single-ended cavity, reducing losses on cavity mirrors, and using homodyne detection), increasing the cavity finesse by  3.5,  and implementing a two-probe laser technique that reduced the relative frequency noise between the probe laser and the empty cavity from 16 to 25 dB relative to the projection noise  \cite{a_CoxIEEE}.  See the Supplemental Material \cite{Note1}.

It is physically reasonable to expect that the majority of the atoms participate in a single multipartite entangled state.  The entanglement depth, or we believe more appropriately the ``entanglement breadth" $\zeta$, quantifies the minimum number of atoms that provably particpate in a multipartite entangled state, no matter how weakly \cite{Sorensen2001,a_VladanHeraldedPhoton}.   We find the largest breadth $\zeta= 400(120)$ atoms at squeezing $S^{-1}$=15~dB, but at the largest squeezing we find  $\zeta$=170(30) atoms.

Applying real-time feedback based on the outcome of joint measurements may allow for new applications in both quantum information technology and precision measurement.  For instance, the utility of highly spin-squeezed states suffers from the fact that the state lives on a sphere, causing the backaction spin projection to couple into the measured spin projection $J_z$ if the state is rotated too far from the equator. In clock applications, this results in needing to reduce the Ramsey phase evolution time such that the net enhancement in clock precision is far from approaching the Heisenberg limit \cite{Andre2004}. It was recently proposed that joint measurement and feedback similar to that used here would allow one to actively measure and steer the backaction noise out of the measured spin projection and would thus allow enhancements in precision approaching the Heisenberg limit \cite{a_sorensen_FB}.  With improved atom-cavity coupling (e.g., higher finesse and smaller mode waist size), even greater amounts of squeezing than that reported here can be achieved in principle \cite{a_SSPRA}. However, it will be critical to consider current limiting effects such as optomechanical ringing and time-varying couplings between measurements due to atomic motion in order to achieve significant improvements.  Having now shown that large enhancements in phase resolution using entanglement are achievable in real systems that are compatible with state-of-the-art precision measurements, the next steps may include application to matter-wave interferometers \cite{a_OberthalerNumberSqueezing}  and optical lattice clocks \cite{2015arXiv150602297N}.

\begin{acknowledgments}
We thank Elissa Picozzi for construction of the homodyne detector and Matthew A. Norcia for helpful discussions. This work is supported by NIST, DARPA QuASAR, ARO, and by the National Science Foundation under Grant No. 1125844.  K. C. C. acknowledges support from the NDSEG fellowship.
\end{acknowledgments}

\nocite{Note1,WinelandParam,WignerRef,CBW12}

\bibliographystyle{apsrev4-1}
\bibliography{final}

%%%%%%%%%%%%%%%%%%%%%%%%%%%%%%%%%%%%%%%%%%%%%%%%%%%%%%%%%
%%%%%%%%%%%%%%%%%%%%%%%%%%%%%%%%%%%%%%%%%%%%%%%%%%%%%%%%%
%%%%%%%%%%%%%%%%%%%%%%%%%%%%%%%%%%%%%%%%%%%%%%%%%%%%%%%%%
%%%%%%%%%%%%%%%%%%%	 SUPPLEMENTARY MATERIALS %%%%%%%%%%%%%%%%%%%%%%%
%%%%%%%%%%%%%%%%%%%%%%%%%%%%%%%%%%%%%%%%%%%%%%%%%%%%%%%%%
%%%%%%%%%%%%%%%%%%%%%%%%%%%%%%%%%%%%%%%%%%%%%%%%%%%%%%%%%
%%%%%%%%%%%%%%%%%%%%%%%%%%%%%%%%%%%%%%%%%%%%%%%%%%%%%%%%%

\clearpage

%%%%%%%%%% Merge with supplemental materials %%%%%%%%%%
%%%%%%%%%% Prefix a "S" to all equations, figures, tables and reset the counter %%%%%%%%%%
\setcounter{equation}{0}
\setcounter{figure}{0}
\setcounter{table}{0}
\setcounter{page}{1}
\makeatletter
\renewcommand{\theequation}{S\arabic{equation}}
\renewcommand{\thefigure}{S\arabic{figure}}
%%%%%%%%%% Prefix a "S" to all equations, figures, tables and reset the counter %%%%%%%%%%

\begin{center}\Large{\textbf{Supplementary Material}}\end{center}

\section{Experimental Details}

Several technological improvements from previous experiments were key to achieving the results of this paper.  Full experimental diagrams of the optics and electronics are shown in Fig. \ref{FullOptical} and Fig. \ref{FullElectronic} respectively in an attempt to highlight differences from previous work, and a general procedure is described.

\subsection{Atomic State Preparation}
The atoms are loaded from a magneto-optical trap (MOT) whose loading time sets the experimental repetition rate of 1 second.  Polarization gradient cooling is then used to cool the atoms to 10~$\mu$K and to load them into 823~nm optical lattice sites spanning approximately 1~mm along the cavity axis.  A bias magnetic field along the cavity axis of 1.1 Gauss sets the quantization axis.  Optical pumping beams are used to polarize the atoms with $>90$\% probability into the $\ket{\downarrow}=\ket{5^2S_{1/2}, F=1, M_F = 1}$ ground state. After optical pumping, the MOT beams are applied once more to clear any remaining atoms from the $F=2$ manifold.  Atoms not in $\ket{\downarrow}$ are not rotated by the microwaves into the measured  $\ket{\uparrow}$ state due to the Zeeman splitting between states. As a result, they do not contribute to the experiment.  Lastly, to account for inhomogeneous coupling of the atomic ensemble to the probe mode, the reported atom numbers $N_\uparrow$, coupling $g$, and cooperativity parameter $\mathcal{C}$, in both the main text and here, are effective values as described in Refs.~\cite{CBS11,a_SSPRA,Squeezing_Bohnet_2014,2015arXiv151001021H}.  Neglecting small corrections for radial inhomogeneity, the total atom number $N_{\uparrow,tot}$ in the lattice in state $\ket{\uparrow}$, the single-atom Rabi frequency $2 g_0$ at an antinode of the probe mode, and the cooperativity parameter $C_0$ at an antinode of the probe mode are related to the effective quantities by $N_\uparrow \approx \frac{2}{3}N_{\uparrow,tot}$, $g^2\approx \frac{3}{4} g_0^2$, and $\mathcal{C}\approx \frac{3}{4}C_0$, respectively.

\subsection{Science Cavity and Lattice}

The optical cavity parameters are given in Table~\ref{tab:CavityParams}.  Compared to previous work \cite{Squeezing_Bohnet_2014,CBS11}, the cavity finesse has been increased by a factor of 3.5, or equivalently the cavity power decay linewidth $\kappa$ is smaller by the same factor.  The cavity is now primarily transmissive at a single end, \textit{i.e.} the input mirror's transmission coupling rate $\kappa_1$ is much greater than the output mirror's transmission coupling rate $\kappa_2$.  As a result, measurement of the probe light in reflection from the cavity captures nearly all of the information transmitted out of the cavity mode, with effective quantum efficiency now of $\kappa_1/\kappa=0.83(3)$ compared to the previous effective quantum efficiency of 0.23 in reflection alone. This eliminates the need for a second detection system for the transmitted cavity light.

The cavity's frequency is actively stabilized to the frequency of the 823~nm optical lattice laser.  This is achieved by a Pound-Drever-Hall (PDH) frequency servo that feeds back to piezos to control the cavity length. The bandwidth of the servo is about $1.5$~kHz.  The lattice laser is frequency stabilized to an independent transfer cavity using PDH detection with servo bandwidth of 1~MHz.  The lattice laser's frequency is tuned relative to the transfer cavity by using a high frequency phase modulator to place 5 to 8 GHz sidebands on the lattice laser light probing the cavity.  The optical frequency of a first order sideband is locked to the transfer cavity, such that tuning the modulation frequency then allows the lattice laser's frequency to be tuned continuously over several GHz.  The microwave voltage-controlled oscillator (VCO) that provides the modulation is phase-locked to a DDS that is controlled by the data acquisition computer for ease of tuning. Finally, the frequency of the transfer cavity is stabilized with 1.5~kHz bandwidth by PDH probing of another longitudinal mode using a 795~nm laser that is stabilized using Doppler-free FM spectroscopy to the D1 transition in $^{87}$Rb.  795~nm light is used simply for historical reasons.

\begin{figure}[htb]
\centering
\includegraphics[width=3.375in]{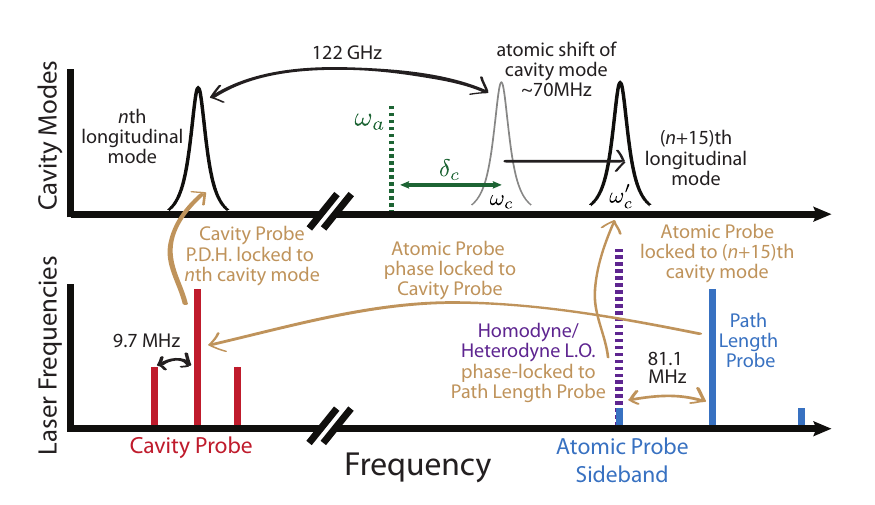}
\caption{Experimental frequency diagram.  Relevant frequencies described in the text are shown along with the locking scheme of the atomic (blue) and cavity (red) probe lasers.  The two longitudinal resonances of the cavity that these two lasers probe are separated by 122~GHz and shown on the upper graph.  The unshifted $n+$15th cavity mode at $\omega_c$ is detuned $\delta_c$ blue from the atomic resonance $\omega_a$.  The presence of atoms in $\up$ typically shift this cavity mode by approximately 70~MHz, to $\omega_c'$.  The homodyne local oscillator beam is shown in purple (dashed), and feedback stabilization steps are shown as gold arrows with descriptions.}  
\label{FD}
\end{figure}

\renewcommand*{\thefootnote}{\fnsymbol{footnote}}
\begin{table}[hbt]
    \centering
    \begin{tabular}{p{.63\linewidth} rp{.34\linewidth}}
        \multicolumn{2}{c}{\textbf{Cavity Parameters (probe $\lambda = 780$~nm)}}\\\hline
        Single-atom cooperativity $\mathcal{C} = \frac{4 g^2}{\kappa\Gamma}$\vspace{1mm} & 0.044(6) \\
        Single-atom vacuum Rabi splitting  $g$ & $2 \pi \times 0.44(3)$ MHz\\
        Input coupling $\kappa_1$ & $2\pi \times 2.60(5)$ MHz\\
        Output coupling $\kappa_2$ & $2 \pi \times 0.17(1)$ MHz \\
        Internal losses $\kappa_L$ & $2 \pi \times 0.38(8)$ MHz\\
        Linewidth $\kappa$ & $2 \pi \times 3.15(10)$ MHz \\
        Dressed-cavity linewidth\footnote{$N = 4 \times 10^{5}$ atoms, $\delta_c = 400$ MHz.} $\kappa'$ & $2 \pi \times 3.6(1)$ MHz \\
        Dressed-cavity linewidth\footnote{$N = 0.5 \times 10^{5}$ atoms, $\delta_c = 400$ MHz.} $\kappa'$ & $2 \pi \times 3.2(1)$ MHz\\
        Q.E. due to internal losses  $\kappa_1/\kappa$ & 0.83(3) \\
        Finesse  & 2532(80)\\
        Free spectral range & 8.105(5) GHz \\
        Frequency difference TEM$_{00}$-TEM$_{10}$  & 2.290(5) GHz \\
        TEM$_{00}$ waist size $w_0$ & 70(1) $\mu$m \\
        Cavity length & 1.849(1) cm \\
        Mirror radius of curvature & 4.999(5) cm \\\hline
        \multicolumn{2}{c}{\rule{0pt}{1.0em}\textbf{Cavity Parameters (lattice $\lambda = 823$~nm)}}\\\hline
        Input coupling $\kappa_1$          & $2 \pi \times 4.40(10)$ MHz \\
        Output coupling $\kappa_2$          & $2 \pi \times 0.23(1)$ MHz \\
        Linewidth           &  $2\pi \times 5.8(6)$ MHz   \\
        Finesse             & $1400(150)$ \\
        Trap depth          & 115 $\mu$K \\
        Circulating power $P_{circ}$   & 0.30(3) W \\
        Power Buildup ($P_{circ}/P_{inc}$) &800(130)\\
        Axial trap frequency    & 181(20) kHz  \\
        Radial trap frequency   & 900(50) Hz \\
        TEM$_{00}$ waist size $w_0$ & 71(1) $\mu$m
    \end{tabular}
    \caption{Relevant cavity parameters at the atomic and cavity probe laser wavelength $\lambda = 780$~nm and at the lattice laser wavelength $\lambda= 823$ nm.  The symmetric, standing wave cavity's mirror transmission coefficients, $T_1$ on the probed end (1) and $T_2$ on the closed end (2), are expressed in terms of coupling rates $\kappa_{1,2}= T_{1,2}\times (\textrm{free spectral range})$.  The atomic decay linewidth of $\ket{e}$ is $\Gamma= 2 \pi \times 6.07$~MHz.  The dressed cavity linewidths $\kappa'$ include broadening of the cavity resonance at $\omega_c'$ due to spontaneous scattering from the atoms.}
    \label{tab:CavityParams}
\end{table}

\begin{figure*}[htb]
\centering
\includegraphics[width=5.5in]{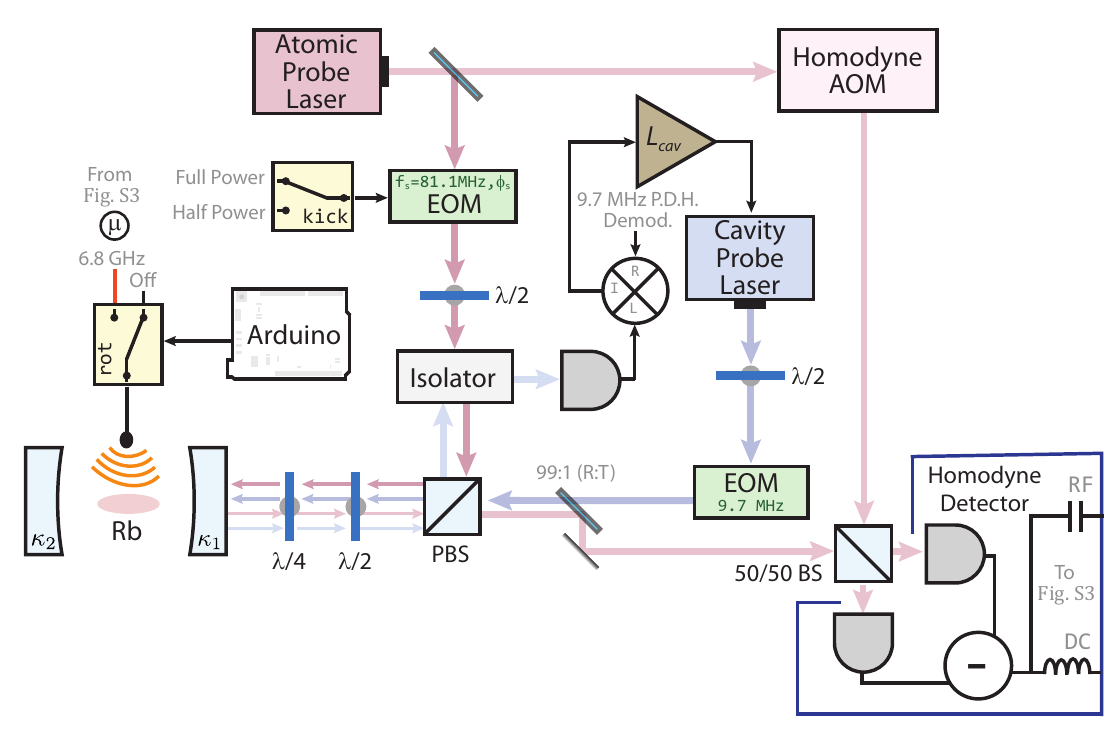}
\caption{Optical block diagram.  The resonance frequency of the optical cavity $\omega_c'$ is detected using homodyne detection of the atomic probe laser (red).  Homodyne detection is performed on an $f_s = 81.1$~MHz sideband on the atomic probe laser.  This sideband can be applied at half power by the ``kick'' switch to provide an extra impulsive kick to the atoms in order to cancel optomechanical ringing (described in Section \ref{subsect:optomech}).  The carrier of the atomic probe laser is detected in heterodyne (\texttt{RF} port) to provide a path length reference (see Fig. \ref{FullElectronic}) for stabilizing the homodyne detection phase.  The cavity probe laser (blue) is P.D.H. locked, via the $\Lcav$ loop filter, to another longitudinal mode of the optical cavity, unshifted by atoms, and provides stabilization of the atomic probe laser's frequency to the cavity frequency.  The atomic probe and cavity probe are separated optically via polarization.  Real-time feedback is applied using an Arduino microcontroller that controls the sign and duration of 6.8~GHz $\mu$-wave pulses.  More details are given in Fig. \ref{FullElectronic}.}
\label{FullOptical}
\end{figure*}

\begin{figure*}[htb]
\centering
\includegraphics[width = 5.5in]{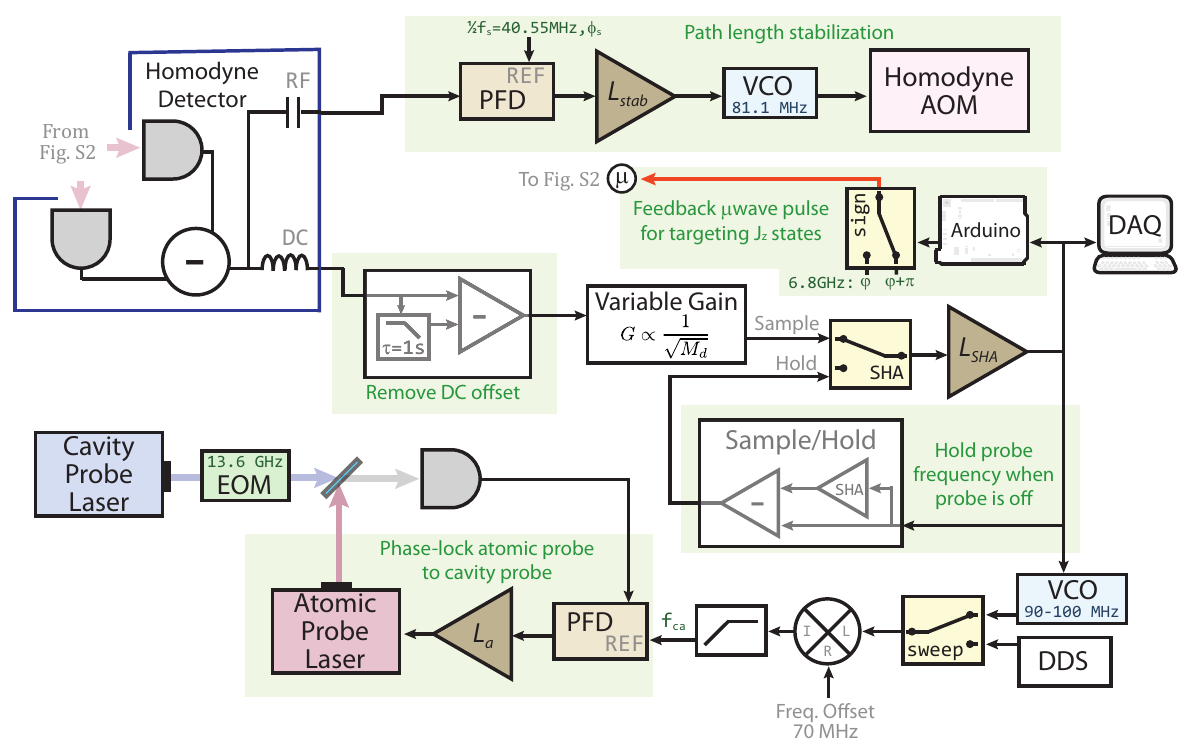}
\caption{Electronic block diagram. The homodyne detection phase is stabilized by detecting the carrier of the atomic probe beam with the signal appearing at $81.1$~MHz at the \texttt{RF} port.  The phase of this signal is locked to a DDS frequency reference by applying feedback through the $\Lstab$ loop filter to a VCO controlling the homodyne AOM. The homodyne difference signal (\texttt{DIFF}) is used to stabilize the atomic probe laser to the atom-shifted cavity mode at $\omega_c'$.  The  signal is high-pass filtered at 1~Hz to remove slowly drifting DC offsets and then passed through a variable gain amplifier (used to maintain constant loop gain as $M_i$ is varied) before entering the loop filter $\Lsha$.   The output of $\Lsha$ is used to control a VCO which provides a phase reference to a phase lock between the atomic probe laser and the cavity probe laser using loop filter $L_a$. The cavity frequency $\omega_c'$ is detected by sampling the output of $\Lsha$.  When the atomic probe is off, a sample and hold circuit is used to hold the output of the loop filter.  A separate synthesizer (DDS) can be used to perform sweeps of the atomic probe.  Real-time feedback is applied by the Arduino based on the sampled output of $\Lsha$.  The Arduino can control the sign of the feedback by switching (\texttt{sign}) between two 6.8~GHz sources that are $180^\circ$ out of phase. }
\label{FullElectronic}
\end{figure*}

\subsection{Relative Frequency Noise Between Cavity and Probe}
\label{stabiliz}
Relative frequency noise between the atomic probe laser (200~kHz FWHM nominal linewidth ECDL laser) and the empty cavity hinders our ability to determine the atomic-induced shift $\omega_c'-\omega_c$ and hence the atomic populations that constitute our joint measurement of spin projection $J_z$.  To remove this, an improved stabilization scheme of the atomic probe laser has been implemented similar to that described in \cite{a_CoxIEEE}, described below and shown in Fig. \ref{FD}.  

A second 200 kHz FWHM  laser at 780~nm, called the cavity probe laser, is PDH locked with servo bandwidth 800~kHz  to a longitudinal mode of the science cavity that is 122~GHz (15 free spectral ranges) away from resonance with the atomic transition $\up$ to $\ket{e}$. The large detuning means that even at much higher circulating powers inside the cavity, the cavity probe produces sufficiently small atomic dephasing and spontaneous emission.  Frequency noise on the science cavity is thereby imposed on the cavity probe laser for spectral noise at frequencies below the unity gain frequency of the servo. Conversely, the original frequency noise of the cavity probe laser is also reduced relative to that of the empty cavity. 

The cavity probe light is circularly polarized $\sigma^-$, opposite to that of the primary or atomic probe which is $\sigma^+$ polarized.  This allows the reflected cavity probe light to be polarization separated from the atomic probe after probing the cavity.  An avalanche photodiode (Hamamatsu S2381, gain $\approx 150$) is used to detect the PDH signal generated by typically 50~nW of total optical probe power.  To maximize the signal-to-noise for a given amount of circulating cavity probe power in the cavity, the PDH signal is derived by phase modulating the cavity probe light at frequency 9.7~MHz $\gg(\kappa/2)/2\pi$ so that the phase modulation sidebands do not enter the cavity.  

To link the cavity probe laser to the atomic probe laser, approximately 2~mW of the cavity probe laser's light is phase modulated at $f_m = 13.6$~GHz.  The modulation frequency is derived from a low-phase noise microwave source \cite{CBW12}.  The atomic probe laser is then phase-locked to a 9th order sideband at a total offset of $9\times f_m=122.4$~GHz from the cavity probe frequency using loop filter $\Lcav$. The heterodyne signal between the sideband and the atomic probe appears in the rf spectrum at 500 to 700 MHz.  This heterodyne beat note is phase-locked to an rf voltage-controlled oscillator (VCO) with center frequency $f_{ca}$, and with servo bandwidth of 2~MHz.  By tuning the VCO frequency $f_{ca}$, we can thereby tune the atomic probe laser relative to the cavity, while frequency noise on the atomic probe is now common-mode with the cavity. 

With this scheme, the measured frequency noise floor $\Delta\omega_c$ for probing the empty cavity resonance $\omega_c$  is approximately $25$~dB below the  typical quantum projection noise induced fluctuations $\Delta\omega_c'/2\pi \approx 100$~kHz of $\omega_c'$, sufficiently small for the work presented.

\subsection{Atomic Probe}
The atomic probe is used to determine the shifted cavity mode frequency $\omega_c'- \omega_c$.  The cavity probe is $\sigma^+$ polarized to take full advantage of the cycling transition for strong light-atom coupling as well as to avoid Raman transitions to other ground states caused by spontaneous emission \cite{a_SSPRA}.  The circular polarization also facilitates easy separation of the cavity probe light reflected from the cavity for sending to a homodyne detector.  

Near resonance, the reflected $q_r$ quadrature response of the field is directly related to the incident field $i_i \propto \sqrt{M_i}$  and the detuning $\delta_p$ between the probe light and the cavity resonance by \cite{a_SSPRA}: 

\begin{equation}
\frac{q_r}{i_i} = \frac{4 \delta_p}{\kappa}\left( \frac{\kappa_1}{\kappa}\right)\left(\frac{\left(\kappa/\kappa'\right)^2}{1+ \left(\frac{\sqrt{N_{\uparrow}} 2 g }{2 \delta_c'}\right)^2}\right)
\end{equation}

\noindent where $\delta_c' = \omega_c'-\omega_a$; $\omega_a$ is the optical atomic transition frequency, and the dressed cavity linewidth is $\kappa' = (\kappa + \Gamma \left(\sqrt{N_{\uparrow}} g/ \delta_c'\right)^2)/(1 +\left(\sqrt{N_\uparrow} g/ \delta_c'\right)^2)$.  Homodyne measurements of $q_r$ allow us to determine the detuning of the probe from $\omega_c'$.

\subsubsection{Homodyne Phase Stabilization}

Homodyne detection requires stabilization of the relative path length  between the homodyne reference path and the probe path, as well as removing other sources of relative phase noise. To achieve this, the homodyne reference light derived from the same laser is shifted up in frequency by an acousto-optic modulator (AOM) driven by a VCO with nominal frequency $f_h = 81.1$~MHz.  The atomic probe light is  weakly phase modulated at fixed frequency $f_s = 81.1$~MHz.  The lower sideband  is tuned close to resonance with $\omega_c'$ while the much stronger carrier component is 81.1~MHz off resonance from the cavity. The strong carrier component primarily reflects off of the cavity without creating any additional light circulating inside of the cavity--important for avoiding dephasing and spontaneous emission from this frequency component.  

The carrier component acts as a phase reference for stabilizing the homodyne detection phase, and it appears on the homodyne detector as a signal at $f_s$.  The signal is separately amplified from the DC signal by AC coupling the homodyne detector's signal to a high frequency transimpedance amplifier AD8015 (\texttt{RF} port). The \texttt{RF} port of the homodyne detector is sensitive to frequencies above $5$~MHz while the DC difference and sum ports (not shown) of the detector have a bandwidth $1.5$~MHz. The two detector ports are balanced well enough that $\frac{P_{DIFF}}{P_{SUM}} < 7 \times 10^{-3}$, where $P_{DIFF}$ and $P_{SUM}$ are the difference and sum of the two powers detected on the two photodiodes comprising the homodyne detector. An additional electronic relative gain adjustment between the two photodiodes allows cancellation of power noise on the homodyne reference by typically $<3\times10^{-4}.$

The carrier/homodyne reference beat note is then phase-locked to a stable DDS reference frequency at $f_s$.  The phase of this reference frequency $\phi_s$ sets the quadrature of detection in homodyne and is under the control of the data acquisition computer.  The phase lock is implemented with 50 kHz bandwidth and is achieved by feedback to the VCO that drives the homodyne frequency shifting AOM at $f_h$.  This feedback loop works to continuously readjust the homodyne reference's phase to compensate for relative path length noise and other relative phase noise so that the $q$ quadrature of the light reflected from the cavity appears at the difference port of the homodyne detector. The rms noise in this phase lock is low enough to resolve the cavity frequency with precision at least 28 dB below quantum projection noise.

\subsubsection{Locking of Atomic Probe to Cavity}

We actively feedback to lock the atomic probe's sideband to $\omega_c'$.  This improves the dynamic range of the detection system, removes sensitivity to scale-factor noise, creates more consistent optomechanical effects, and removes nonlinearities associated with the dispersive error signal.  The error signal is the detected $q$ quadrature of the atomic probe's lower sideband as measured in homodyne at the difference port. The signal is a dispersive feature with a zero crossing appearing as the atomic probe laser's frequency is swept through resonance with $\omega_c$ (or $\omega_c'$) \cite{a_SSPRA} .

During each measurement window of $\omega_c'$ the atomic probe's lower sideband is turned on for approximately $40~\mu$s, and the DC homodyne signal is used to actively lock the sideband's frequency to $\omega_c'.$  This is achieved by feedback to the VCO that provides the frequency reference $f_{ca}$ to which the cavity/atomic probe beat note is phase-locked.  The phase-locking is achieved by adjusting the atomic probe laser's frequency via the loop filter $\Lsha$.  The characteristic settling time of the servo is $1~\mu$s for a unity gain frequency of 160~kHz.  In order to record $\omega_c'$, the output of the $\Lsha$ loop filter that sets the VCO control voltage is directly sampled at 2.5~MHz by the data acquisition computer (DAQ).  

Since the atomic probe lower sideband is turned off between measurements, the atomic probe laser's frequency must be held fixed using the sample-and-hold circuit as shown in Fig. \ref{FullElectronic}.   When the atomic probe sideband is turned on, the circuit samples the loop filter voltage provided to the VCO that provides $f_{ca}$.  When the sideband is turned off, the circuit holds the output voltage of the loop filter so that $f_{ca}$ is held at its previous value.

Trial-to-trial fluctuations in atom number are significantly larger than fluctuations due to projection noise. This increases the range over which the probe laser must slew its frequency to align the lower sideband with $\omega_c'$ during the first $\NuO$ premeasurement.  To reduce this initial offset,  a ``pre-centering'' measurement is performed 1.5 ms before each experimental squeezing trial: a $\pi/2$ microwave pulse rotates the atoms to a superposition of $\up$ and $\down$ and the lower sideband is centered by the feedback loop at $\omega_c'$.  The atomic probe frequency is then held, and the probe light is switched off.  The atoms are then optically pumped back to $\ket{\downarrow}$ for the actual spin squeezing measurements described in the main text.

We often wish to scan the power in the atomic probe lower sideband (quantified by the number of incident probe photons in a single measurement window $M_i$) to look at variation in measurement noise. This is accomplished by changing the rf power supplied to the EOM at $f_s$ to modify the phase modulation index.  For reference, a typical sideband/carrier ratio for $M_i = 36500$ incident photons is $0.004$.  Thus, the carrier power and hence the open loop gain of the path length phase stabilization for homodyne detection is relatively unaffected as we vary $M_i$.

In contrast, the open loop amplitude gain of the lower sideband to cavity lock scales as $\sqrt{M_i}$.  To compensate, a variable gain amplifier (VGA; Analog Devices AD8337) is inserted after the homodyne detector.  When the data acquisition computer changes the rf power that sets $M_i$, it also  simultaneously scales the VGA's gain to keep the net loop gain fixed.   DC offsets in the homodyne difference port are problematic when the gain is scaled and are therefore removed using a low pass filter ($\tau=1$~s) and differential amplifier that essentially make a low bandwidth measurement of the DC offset that is then subtracted from the fast $40~\mu$s measurement windows.  

With this approach, we achieve a very large dynamic range from $M_i = 150$ to $M_i = 3\times 10^5$.  When $M_i \lesssim 100$ in a 40~$\mu$s window, the average number of detected photons within the servo's time scale of $1~\mu$s approaches unity.  The photon shot-noise then imposes rms fluctuations on the atomic probe's frequency that are comparable to the cavity half-linewidth, leading to a reduction in fundamental signal to noise for estimating $\omega_c'$.

Lastly, for diagnostic reasons, it is often useful to do broad sweeps of the lower sideband's frequency across the cavity resonance frequency.  To accomplish this, the atomic probe laser's beat note with the cavity probe laser can be phase-locked to  a direct digital synthesizer (DDS) source that provides the reference frequency $f_{ca}$ in place of the usual VCO.  The DDS frequency can be phase-coherently swept at programmable rate and range, accomplishing the desired sweep of the atomic probe frequency.

\subsubsection{Calibration of Incident Photon Number $M_i$}
 The number of incident photons on the cavity $M_i$ is determined from the homodyne signal and measured quantum efficiencies.  The locking scheme used for homodyne detection allows precise control of the relative phase between the homodyne reference beam and the atomic probe sideband by tuning the phase $\phi_s$.  Experimentally, when the atomic probe sideband is off resonance from the cavity and one scans $\phi_s$ over $2 \pi$, a sinusoidal interference fringe is observed in the homodyne difference port. The size of this fringe and the independently measured total power in the homodyne reference beam ($130$ $\mu$W typical) are used to determine the rate $R_i$ of incident photons in the atomic probe lower sideband, coupled to the cavity, that would have been required to produce the observed fringe. The number of incident photons is $M_i=R_i\times 40~\mu$s.  

Physically, this means $M_i$ can be understood as the average number of photons in the atomic probe lower sideband crossing an imaginary plane directly in front of the cavity input mirror, counting only those that are spatially mode matched to the cavity TEM$_{00}$ mode, and integrated into a 40 $\mu$s window.  The uncertainty in the absolute calibration $M_i$ is approximately $25\%$ due to uncertainty in the spatial mode-matching of the incident atomic probe beam and the homodyne reference beam.  This uncertainty leads to uncertainty in the prediction of contrast lost in Fig. 3(b) of the main text, but does not lead to any uncertainty in the amount of squeezing or the experimental quantum efficiency $\Qcompoundexpected$ to be discussed in Section~\ref{ss:qe}.

\subsubsection{Quantum Efficiency}
\label{ss:qe}

To determine the probe detuning $\delta_p$, we estimate the ratio $q_r/i_i$ from the detected fields $q_d/i_d$. Vacuum or photon shot noise that appears in the detection of the $q_d$ quadrature limits the resolution on our ability to determine $\omega_c'.$   We express the noise in the ratio as

\begin{equation}
\frac{(\Delta q_d)^2}{i_d^2}=\frac{1}{4 M_i \Qcompoundexpected}+(\Delta q )^2,
\end{equation}

\noindent where the one-window quantum efficiency $\Qcompoundexpected$ includes fundamental losses of signal to noise resulting from both photon losses and technical noise floors shown in Table \ref{tab:qe}.  

The additional term $(\Delta q)^2 = f+r M_i^n $ represents  noise contributions from the technical noise floor $f$ associated with residual frequency noise on the atomic probe laser relative to the cavity mode frequency, and noise from optomechanical ringing $r$, which we model with an arbitrary $n$th-order polynomial scaling with $n\neq-1$ .   These noise sources have different scalings with $M_i$ than the fundamental quantum noise (first term).

We define a new effective quantum efficiency $\Qcompound$ which includes the effects of the technical noise floor and optomechanics and write the noise in homodyne detection as 

\begin{equation}
\frac{(\Delta q_d)^2}{i_d^2} = \frac{1}{4 M_i \Qcompound}
\end{equation}
where $\Qcompound$ is given by
\begin{equation}
\Qcompound = \frac{\Qcompoundexpected}{1+4 M_i \Qcompoundexpected (f + r M_i^n)  }.
\end{equation}
This effective quantum efficiency provides a useful figure of merit for the experiment and can be compared to measurements of the increase in area of the Bloch vector's noise distribution (discussed in Section \ref{sec:qe}).

\begin{table}[hbt]
    \centering
    \begin{tabular}{l l}
        \textbf{Source} & \textbf{Q}  \\\hline
        \rule{0pt}{1.0em}Path efficiency, $Q_{path}$            & 0.75(3) \\
        Cavity-mode/homodyne overlap, $Q_{o}$  \qquad           & 0.95(3) \\
        QE of cavity $(\kappa_1 / \kappa)$, $Q_{cav}$           & 0.83(3) \\
        Technical noise from detector, $\Qelectronic$           & 0.86(1) \\
        Detector QE, $\Qsensitivity$                            & 0.86(2) \\
        Probe turn-on time, $\Qdeadtime$                         & 0.86(1) \\\hline
        Total, $\Qcompoundexpected$  & 0.37(5)
    \end{tabular}
    \caption{Quantum efficiency summary table. Quantum efficiency losses come from sources of signal loss and added noise floors. $\Qdeadtime$ comes from finite laser turn-on times and ringing-cancelling ``kicks'' (see Sec. \ref{subsect:optomech}) during which the probe is on but we do not collect information. The total quantum efficiency $\Qcompoundexpected=0.37(5)$ is the product of all the measured contributions.}
    \label{tab:qe}
\end{table}

\subsubsection{Limits to Noise Reduction, Optomechanics}
\label{subsect:optomech}

\begin{figure}[htb]
\centering
\includegraphics[width=3.375in]{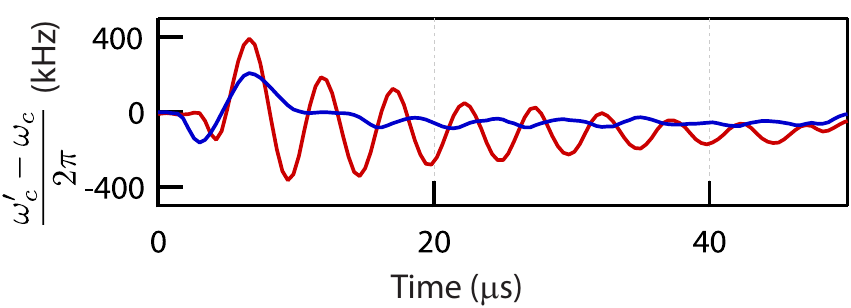}
\caption{Probe induced oscillations partially cancelled by a staggered turn-on sequence. The oscillations are fully present with no kick (red, 43 traces averaged) during a 40 $\mu$s measurement, but greatly reduced by a half-power 2.5 $\mu$s kick (blue, 30 traces averaged). The 2.5 $\mu$s kick length corresponds to a quarter of the axial trap oscillation period. There is an 80 MHz offset subtracted from the vertical axis.}
\label{RingingFig}
\end{figure}

The primary limitation to noise reduction $R$ is currently set by optomechanical effects from the probing light.  Due to the incommensurate probing and trapping potentials, when the probe light is turned on, the atoms are given an impulse that drives axial oscillations in the trap.  Additionally, the minimum of the trapping potential moves in space.  This ringing effect is shown in Fig. \ref{RingingFig} (red) over the 40 $\mu$s probing period.

 To partially cancel the optomechanical ringing, we employed a 2.5~$\mu$s half-power turn-on sequence of the probe laser.  The initial half power turn-on induces ringing, while the second, full-power turn-on (applied one quarter of an axial oscillation period later) coherently zeroes the initial axial ringing such that the atoms come to rest at the new trap minimum.  As shown in Fig. \ref{RingingFig}, this technique significantly reduced the amount of ringing but only somewhat improved the optimal squeezing in Fig.~3 by an estimated 0.6 dB.  Mitigation of optomechanical effects will present a challenge for future experiments aimed at generating even more spin squeezing.  Tighter trapping or homogeneous coupling of atoms to the atomic probe could be avenues toward reducing optomechanical effects.  

\section{Background Contrast Correction}

The spin squeezing values presented in the main text were calculated using the slightly modified relationship $S = R C_{BG} /C^2$ described in Ref. \cite{WinelandParam}.  Here $C_{BG} = 0.96$ is the background contrast as determined from measurements of the contrast at $M_i =0$ probe photons.  For clarity of presentation we approximated $C_{BG}=1$ for the expressions in the main text.  Using the more exact formula represents a small 0.2~dB improvement in the reported squeezing compared to what would be calculated with $S = R /C^2$.

%parameter $S \equiv (\Delta \theta / \Delta \theta_{\text{SQL}})^2 = R /C^2$ presented in the main text is corrected for a finite background contrast $C_{BG}=0.96$.  The background contrast loss is believed to be due to imperfect spin echo of inhomogeneous broadening from the optical lattice and the cavity probe beam discussed in Section \ref{stabiliz} and is determined by measuring the contrast at $M_i = 0$.  We calculate squeezing using the formula $S = R C_{BG} /C^2$ as decribed in Ref [??].  This leads to a small 0.2~dB increase in the squeezing $S^{-1}$ relative to the formula with no correction for $C_{BG}$.

%Given this imperfect background contrast, the properly adjusted $S$ must be re-normalized with respect to the SQL of the shortened Bloch vector $\theta_{\text{SQL}} =1/\sqrt{N C_{BG}}$.  Since $C$ already includes $C_{BG}$ (essentially over-counting the effect background contrast) this correction is done by adding a factor of $C_{BG}$ to the numerator of $S$, $S = R C_{BG} /C^2$.  This leads to a small 0.2~dB increase in the squeezing $S^{-1}$.

\section{Antisqueezing and Area of the Noise Distribution}
\label{sec:qe}

The noise in the backaction (or antisqueezed) quadrature of the squeezed state was measured using the sequence in Fig.~3(c) of the main text, with the variable rotation $\psi$ inserted, and measurement outcomes here labeled $J_{zp}$ and $J_{zf}(\psi)$ for the first and second measurements respectively.  This measurement sequence was used to make the visualization of the squeezed state shown in Fig. 3(e) of the main text.  In order to construct a meaningful probability distribution describing our state, we constructed the normalized probability distribution $P\left(J_{zf}\left(\psi\right)-\cos{\psi}J_{zp}\right)$ for obtaining a differential measurement outcome $J_{zf}(\psi)-\cos{\psi}J_{zp}$.  The weighting of the premeasurement by $\cos{\psi}$ ensures that we only condition the final measurement on the premeasurement to the degree that the two spin projections overlap.  We performed an inverse Radon transform \cite{WignerRef} on the measured $P\left(J_{zf}\left(\psi\right)-\cos{\psi}J_{zp}\right)$, yielding the conditional probability distribution shown in Fig. 3(e) of the main text.

%The noise in the backaction (or antisqueezed) quadrature of the squeezed state was measured using the sequence in Fig.~3(c) of the main text, with the variable rotation $\psi$ inserted.  This measurement sequence was used to make a visualization of the squeezed state shown in Fig. 3(e) of the main text.  In order to reconstruct the probability distribution of the final state, relative to the premeasurement, we define the probability distribution $P(J_z,\psi)$ of obtaining a final measurement value $J_{zf}$ relative to the premeasurement $J_{zp}$ after a particular $\psi$-rotation, $P(J_z,\psi)=P(J_{zf}(\psi)-\cos{\psi}J_{zp})$.  By inserting the cosine, we only condition on the premeasurement when it is in the same quadrature as the final measurement.  To construct the probability distribution shown in Fig. 3 (e) of the main text, we performed an inverse Radon transform \cite{WignerRef} on $P(J_z,\psi)$, yielding the plotted conditional probability distribution $P(J_{zf}-J_{zp}, J_x)$.  

We now consider the magnitude of the noise in the backaction quadrature versus the number of probe photons $M_i$.  We generalize the spin noise reduction to now be a function of $\psi$ as $R(\psi)= \Delta\left(J_{zf}\left(\psi\right)-\cos{\psi}J_{zp}\right)^2/\Delta J_{z,QPN}^2$.  The antisqueezing is defined as $A \equiv R(\pi/2) C_{BG} /C^2$, in direct analogy to the Wineland squeezing parameter, $S = R(0) C_{BG}/C^2$.  The antisqueezing parameter can be interpreted as the noise variance in the azimuthal phase of the Bloch vector relative to the standard quantum limit $A \approx (\Delta \phi / \Delta \phi_{\text{SQL}})^2$, up to the small correction for the background contrast $C_{BG} = 0.96$.  

%the antisqueezing $A$ can be defined in the same way after a $\psi = \pi/2$ rotation, $A \equiv R|_{\psi=\pi/2} C_{BG} /C^2$.  When $A \gg 1$, the antisqueezing parameter can be interpreted as the noise variance in the azimuthal phase of the Bloch vector relative to the standard quantum limit $A \approx (\Delta \phi / \Delta \phi_{\text{SQL}})^2$, up to the small correction for the background contrast $C_{BG}\approx 0.95$.  
%One key improvement for this work is a large increase in experimental quantum efficiency for detecting the atomic probe lower sideband. We consider any loss of signal to noise as an effective loss of quantum efficiency.  The total experimental quantum efficiency is related to the increase in area of the quantum noise distribution on the Bloch sphere and, similarly, to the loss in purity of the quantum state.  Quantum efficiency loss in our experiment comes from optical losses, detector inefficiencies, and technical noise floors.  Table \ref{tab:qe} lists the primary sources of quantum efficiency loss in our experiment, leading to a total quantum efficiency for a single joint measurement, $Q_{exp,pred} = \prod Q = 0.38$\color{red}(??)\color{black}.  The quantum efficiency loss is dominated by the path efficiency from the science cavity to the detector, the technical noise floor of the detector, losses inside the optical cavity, and the quantum efficiency of the heterodyne detector.

\begin{figure}[htb]
\centering
\includegraphics[width=3.375in]{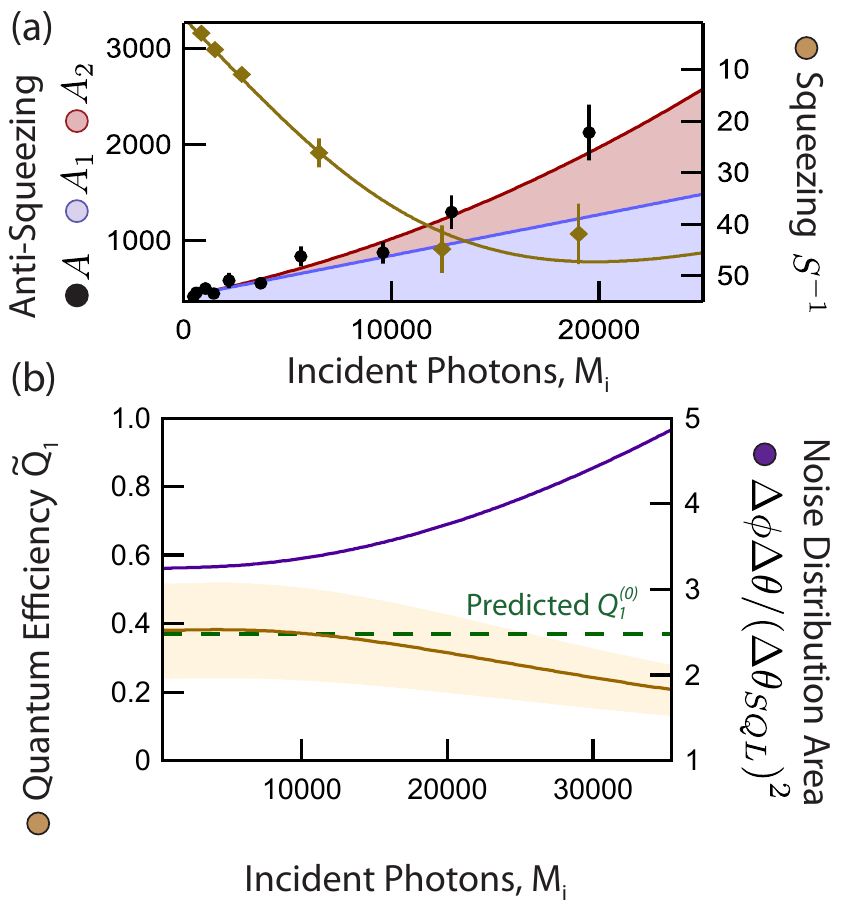}
\caption{(a) The antisqueezing $A$ is plotted versus $M_i$ (black circles).  The linear contribution to  the rise in $A$, $A_1$, is shown in blue and the quadratic contribution $A_2$ in red.  The squeezing (gold diamonds and fit) is plotted on the right axis.  (b)  The area of the noise distribution is calculated from the data in part (a) and plotted in purple.  The measured effective quantum efficiency $\Qarea$ is plotted in gold with an error bar shown as a gold band.  At low $M_i$, $\Qarea$ is consistent with the prediction (green dash) $\Qcompoundexpected$ from Table \ref{tab:qe}.}  %backaction noise variance $(\Delta \phi / \Delta \phi _{\text{SQL}})^2$ is plotted versus $M_d$.  The data is fit (red line) to a model with quantum (linear, blue) and classical (quadratic) scaling.  Spin squeezing taken along with this data is shown in black.  (b) From the data in (a), the quantum efficiency of a joint measurement $Q_{exp}$ (gold) is extrapolated versus $M_d$.  The area of the quantum noise distribution $??$ is plotted in purple.  The quantum efficiency (area) begins to fall (rise) at high $M_d$ due to additional technical noise, primarily from optomechanical ringing.      \color{red}{Change units to angles, $\Delta \theta \Delta \phi$?  .What is really plotted on the right of part (b)?}}
\label{BA}
\end{figure}

The antisqueezing $A$ is plotted in Fig.~\ref{BA} versus $M_i$.  The data is fit to a model that includes three contributions $A = A_{0}+A_{1}M_i+A_{2}M_i^2$.  The quantum backaction should rise linearly with $M_i$ and is therefore parameterized by $A_{1}$.  The contribution of this term to the total backaction is shown by the blue shaded region. Classical intensity noise on the probe laser power circulating inside the cavity (for example) would contribute backaction noise scaling as $M_i^2$.  The classical backaction is therefore parameterized by $A_{2}$, with this classical contribution to the total backaction shown by the red shaded region.
Lastly, the constant term $A_{0}$ is attributed primarily to the projection noise as well as noise in the rotations.

%(red line) is attributed to classical noise in the $\psi$ rotation,

%Lastly, the constant term $A_{0}$ is attributed primarily to the projection noise in the first measurement as well as  noise in the rotation through angle $\psi= \pi/2$.  

Squeezing data $S$ was taken at the same experimental settings (gold points and line  in Fig.~\ref{BA}).  This allows us to infer the angular area of the quantum noise distribution, $\Delta \phi \Delta \theta/\Delta\theta_{\text{SQL}}^2 = \sqrt{S A_{1}/C_{BG}^2}$, shown in purple in Fig. \ref{BA}(b).  

%\textcolor{red}{I think this paragraph needs to be updated to match the discussion of quantum efficiency a little more where Q'exp is defined.    Does this mean that we should move the quantum efficiency section back to here in order to explain what goes into the predicted Qexp?}

The increase of the area of the noise distribution can also be used as an alternate, global measurement of the quantum efficiency $\Qcompound$ in Section \ref{ss:qe}.  Specifically, the total quantum efficiency of the entire measurement sequence is proportional to the square of the increase in the angular area of the noise distribution and can be written $\Qarea=4/(A_{1} S C^2/C_{BG}^2)$. The factor $C^2$ comes from the angular momentum uncertainty relation and accounts for the fact that the SQL increases as the Bloch vector shrinks.  As mentioned in the main text, the factor of four arises due to finite measurement strength and an unused premeasurement. $\Qarea$ as measured by the area of the noise distribution is plotted in Fig. \ref{BA}(b) in gold.  The gold shaded region represents the uncertainty in the extrapolation of $\Qarea$ due to uncertainty in the fit of the antisqueezing data of Fig. \ref{BA}(a).  At low photon number, $\Qarea$ agrees with the predicted value of $\Qcompoundexpected$ from Table \ref{tab:qe}.  At higher photon number, $\Qarea$ begins to rise due to the effective quantum efficiency losses from the technical noise floor and optomechanics discussed in Section \ref{ss:qe}.% $Q'$ \textcolor{red}{Is this Q'exp?} from optomechanics and the technical noise floor.

\section{Real-Time Feedback to a Target $J_z$}
\subsection{Experimental Details}
Real-time feedback to steer the atomic spin projection to a target spin projection $J_z$ is implemented with an Arduino Due microcontroller with an internal clock of 84~MHz.  The microcontroller is programmed to sample the loop filter $\Lsha$ output during the $\NuO$ and $\NdO$ measurement windows, and the sampling rate allows for averaging 18 points in each $40~\mu$s window. The microcontroller then calculates $\Jzp$ from the difference of the two measurement windows and applies the feedback microwave rotation to the atoms through the formula $\thetafb\approx 2\times(\Jztar-\Jzp)/N$.  Fluctuations in $N$ are small enough that it can be taken as a constant.  The microcontroller controls the microwave rotation angle by varying the duration for which the microwaves are applied using  a high speed microwave switch (labelled \texttt{rot} in Fig. \ref{FullElectronic}) with single clock cycle (12 ns) resolution.  The sign of the rotation is controlled by another digital output of the Arduino that toggles a switch denoted \texttt{sign} between two microwave sources that are $180^\circ$ apart in phase, as shown in Fig. \ref{FullElectronic}.  Microwave rotations to accomplish $\pi/2$ and $\pi$ pulses can also be applied independently of the Arduino using digital outputs from the data acquisition computer (not shown), though with less timing resolution.

%\begin{table}[hbt]
%    \centering
%    \begin{tabular}{p{.15\linewidth} p{.8\linewidth}}
%        \textbf{SHA} & \\\hline
%        On  & Use \texttt{HOLD} voltage from the sample/hold amplifier to drive loop filter while the probe is off \\
%        Off & Use homodyne \texttt{DIFF} signal to run the experiment. Also sample loop filter $L_s$ into the sample/hold amplifier during this time.  \\\hline
%        \textbf{sweep} & \\\hline
%        On  & Use DDS to sweep atomic probe freq. \\
%        Off & Lock atomic probe to VCO. \\\hline
%        \textbf{rot} & \\\hline
%        On  & Broadcast 6.8 GHz coherent microwave rotation. \\
%        Off & Microwaves off \\\hline
%        \textbf{kick} & \\\hline
%        On  & Attenuate atomic probe by half during kick \\
%        Off & Probe at full power.
%    \end{tabular}
%    \caption{Truth table for switches in Figs. \ref{FullElectronic} and \ref{FullOptical}}
%    \label{tab:truth}
%\end{table}

%\subsection{degree of squeezing as you tune the feedback gain coefficient...}
\subsection{Limitations to Squeezing with Feedback}
As mentioned in the main text, the primary limitation to deterministic squeezing is noise imposed from microwave rotations.  We estimate these noise sources by performing two additional variations of the measurement sequence of Fig.~2 of the main text, removing either the feedback rotation $\theta_{FB}$ or all rotations.

At the optimal deterministic spin squeezing with feedback, we achieve $R^{-1} = 9.5(4)$~dB.  To estimate the noise added from feedback, we measure conditional spin noise $\Jzf-\Jzp$ in a sequence with no feedback and find $R^{-1} = 12.4(7)$~dB.  Feedback leads to approximately 2.9~dB of added noise.  Next, we perform the sequence with no microwave rotations of any kind, effectively measuring the same spin population $\Nu$ four times.  In this sequence we attain $R^{-1} = 14.0(5)$~dB, 1.6~dB less than the sequence with rotations but no feedback.  This measurement suggests a rotation noise floor due to microwave amplitude and frequency noise that is approximately 17.5~dB below projection noise.  Further, we suspect that rotation noise is also a primary contribution to the additional noise from adding feedback, since certain rotation errors which cancel after two $\pi$ pulses will no longer cancel when feedback is applied.  Improving the precision of microwave rotations remains a major obstacle in working with atomic spin states with extreme phase resolution.

\bibliographystyle{apsrev4-1}
\bibliography{final.bib}

\end{document}